\documentclass[pra,twocolumn,notitlepage,showpacs, superscriptaddress,nofootinbib,amsmath,amsthm,amssymb,UFT8]{revtex4-1}
\usepackage{graphicx} %
\newcommand{\cellcolor}[1]{} %
\usepackage[letterpaper,margin=1in]{geometry}
\usepackage{algorithm}
\usepackage{amsthm, mathtools, mathrsfs, amsfonts, amssymb, amsmath, physics,array}
\usepackage{titlesec}
\usepackage[usenames,dvipsnames]{xcolor} %
\usepackage{tcolorbox}
\usepackage{ragged2e} %

\usepackage{tabularx}
\usepackage{todo}
\usepackage[normalem]{ulem}
\usepackage{booktabs}
\usepackage{mathtools}

\newcounter{protocol}

\definecolor{jpmcbrown}{HTML}{875C3F}
\definecolor{jpmcblue}{HTML}{B0D6ED}
\definecolor{jpmcbluedark}{HTML}{8FBCCF}

\newtcolorbox{mybox}[1][]{
    colback=jpmcbrown!5,
    colframe=jpmcbrown,
    fonttitle=\bfseries,
    title=#1,
    before=\par\noindent\hspace*{\parindent}, %
    after=\par, %
    parbox=false, %
}

\usepackage{xurl}
\usepackage[colorlinks,allcolors=blue]{hyperref}
\usepackage{cleveref}
\definecolor{amaranth}{rgb}{0.9, 0.17, 0.31}

\newcommand{\rev}[1]{#1}
\newcommand{\old}[1]{}

\begin{document}

\title{Applications of Certified Randomness}
\author{Omar Amer}
\author{Shouvanik Chakrabarti}
\author{Kaushik Chakraborty}
\author{Shaltiel Eloul}
\author{Niraj Kumar}
\author{Charles~Lim}
\author{Minzhao Liu}
\author{Pradeep Niroula}
\author{Yash Satsangi}
\author{Ruslan Shaydulin}
\thanks{Corresponding Author.\\Email: \url{ruslan.shaydulin@jpmorgan.com}}
\author{Marco Pistoia}
\thanks{Principal Investigator and Corresponding Author.\\Email: \url{marco.pistoia@jpmorgan.com}}
\affiliation{Global Technology Applied Research, JPMorganChase, New York, NY 10001, USA}

\begin{abstract}
    Certified randomness can be generated with untrusted remote quantum computers using multiple known protocols, one of which has been recently realized experimentally~\cite{jpmc_cr}. Unlike the randomness sources accessible on today's classical computers, the output of these protocols can be certified to be random under certain computational hardness assumptions, with no trust required in the hardware generating the randomness. In this perspective, we explore real-world applications for which the use of certified randomness protocols may lead to improved security and fairness. We identify promising applications in areas including cryptography, differential privacy, financial markets, and blockchain.
    Through this initial exploration, we hope to shed light on potential applications of certified randomness.
\end{abstract}

\maketitle

\section{Introduction}

Randomness is a crucial resource permeating many aspects of everyday life. For example, it is required by cryptographic primitives that ensure the security of online transactions. More generally, uniform and independently sampled randomness is a prerequisite for the fair and trustworthy completion of essential activities such as medical research (randomized clinical trials), financial markets (randomized order execution, allocation of shares through lottery in an oversubscribed initial public offering), risk management and financial regulation (audits), and democratic governance (jury duty selection, draft lottery, census). In fact, the legitimacy and trustworthiness of such activities have in the past been undermined by imperfect or non-verifiable randomness~\cite{starr1997nonrandom,Berinsky2015,Marcondes2019}, increasing the scrutiny of how input randomness is generated~\cite{Garfinkel2020,Jadoul2023}. 

The challenge is compounded by the fact that randomness is difficult to verify. In the classical world, the notion of randomness or entropy denotes a lack of knowledge: some bits are random if an adversary lacks the information necessary to predict them. While the \emph{quality} of randomness can be evaluated using standard tests like NIST SP 800-90~\cite{Barker2015,Turan2018,Barker2024}, it is difficult to verify that the randomness is indeed sampled independently from an adversary's prior knowledge. For example, an attacker may replace the hardware security module generating randomness with a memory stick containing bits that were honestly sampled in advance. The output of such a ``memory stick'' device would look random to the user even though it is completely predictable to the attacker \cite{acin2016certified}.

The substantial demand for randomness in applications makes it necessary today to resort to unverifiable classical solutions, which create vulnerabilities. Hardware security modules accessed via an API~\cite{AWS_KSM_HSM} or the RDSEED instruction built into many CPUs~\cite{IntelRDRAND,AMD_RDRAND} can produce a large amount of randomness on demand. However, these classical approaches have been criticized as difficult to audit and prone to hard-to-detect failures~\cite{Garfinkel2020}. For example, one implementation of RDSEED has been shown to lack robustness guarantees against adversaries who can see a large number of the instruction outputs~\cite{Shrimpton2015}.
Consequently, the Linux operating system does not rely solely on such hardware solutions and combines multiple sources of randomness, such as patterns of memory access, key strokes, etc.~\cite{LinuxRandom} \old{Alternatively, a}\rev{A} small trusted seed \rev{generated in this way} can \rev{then} be expanded using a cryptographically secure pseudorandom number generator (PRNG). However, the output would only be as secure as the random seed, which may be possible to determine by observing the output of \rev{a} PRNG if the PRNG has a backdoor like was the case for the standardized and deployed Dual\_EC\_DRBG~\cite{rsa_backdoor,rsa_backdoor2}.

It has been demonstrated that quantum information protocols can be used to generate
randomness with no dependence on the internal state of the quantum devices used. These protocols~\cite{colbeck2011private,pironio2010random,hensen2015loophole,shalm2015strong,acin2016certified}, beginning from the work of Colbeck and Kent~\cite{colbeck2011private}, rely on the observation that non-communicating devices sharing quantum entanglement can obtain a greater winning probability on certain non-local games than that of unentangled classical devices, with the difference in winning probability characterized by the Bell~\cite{bell2004speakable,brunner2014bell} \rev{or related~\cite{mironowicz2024generalized,joarder2022loophole,nath2024single}} inequalities. Additionally, when \old{Bell}\rev{such} inequalities are violated, the responses of the non-communicating devices must contain genuine new randomness. The primary downside of these protocols is that when accessing the device as a cloud-based application, the client may find it challenging to ensure the lack of communication or other loopholes, making it necessary to trust the remote device or its provider.

To get around this challenge, multiple protocols have recently been proposed for generating certified random numbers using quantum computers under only computational assumptions~\cite{aaronson2023certified,mahadev2022efficient,yamakawa2022verifiable} \rev{and no physical assumptions about the device.}
\old{(see Box 2 for more details). These protocols are modeled by an interactive protocol between a quantum generator and a classical verifier (or client), where the randomness is certified to be genuine if the generator's outputs are accepted by some classical computation. Under some computational assumptions, acceptance by the verifier guarantees that the output of the protocol is secure without needing to trust the generator. These protocols take a small amount of input randomness and expand it into larger amount of output randomness, with the output being independent from the input and thereby unpredictable even to an adversary who knows the input. For example, the protocol proposed by Aaronson and Hung~\cite{aaronson2023certified} only requires that the input randomness is secure against the quantum computer operator, whereas the output has everlasting security. Importantly, the assumptions for such a protocol correspond to the computation time available to the generator, which can be remotely enforced via latency requirements.}\rev{Notably,} this allows certified randomness to be obtained even using a quantum computer accessed remotely over the internet, which has been realized experimentally using a trapped-ion quantum processor~\cite{jpmc_cr}.

Not all applications naturally benefit from certified randomness. For example, randomized algorithms such as Monte-Carlo simulations only require the uniformity of the input randomness and not its unpredictability to an adversary. Such applications do not clearly benefit from the certification of randomness. On the other hand, for many applications, the security, trustworthiness and fairness of randomness are of utmost importance. One popular example is lotteries, which are still commonly decided in an elaborate ceremony involving physical devices and televised live~\cite{cnnPowerball}. 

In this perspective, we propose applications which may benefit from using certified randomness in concrete ways. Our proposed applications fall in three domains. 
First, we present applications to cryptography, where we discuss how \old{quantum randomness expansion protocols}\rev{certified randomness} may be used to immunize backdoored PRNGs and how jointly certifiable randomness may improve the security and reduce the resource requirements of non-interactive zero-knowledge proofs. Additionally, we consider a randomness beacon that uses certified quantum randomness, a primitive that is leveraged by multiple applications. Second, we present applications related to fairness and privacy, where we show that certified randomness may be applied to improve the guarantees offered by differential privacy, as well as the transparency of financial markets. Third, we present the ways in which decentralized applications using blockchain can benefit from certified randomness. We believe formal security analysis of these applications to be a fruitful direction for future research.

\section{Certified Randomness}\label{sec:general_definition}

There have been several proposed protocols for certifiable randomness generation from quantum computers that rely only on computational assumptions~\cite{brakerski2021cryptographic,yamakawa2022verifiable,aaronson2023certified}. Each of these protocols involves an interaction between a classical verifier (randomness consumer) and a quantum prover (randomness generator), where the generator is asked to perform a series of  computational tasks (\rev{which may be chosen} adaptively), and the series of responses is either accepted or rejected by the verifier \rev{based on the outcome of some classical computation. Acceptance by the verifier guarantees that the output of the protocol is random without needing to trust the generator.}

The purpose of the interaction is twofold. Firstly, classical computation is fundamentally deterministic, making it impossible for a classical function to increase the amount of randomness. The challenges posed to the prover must therefore be sufficient to convince the classical verifier that the prover is behaving non-classically, e.g., via a \emph{proof of quantumness}. Furthermore, the protocol must ensure that an accepted transcript of messages includes new randomness within the responses of the quantum prover. This is \old{quantified by requiring}\rev{reflected by the requirement} that the messages from the prover have non-zero entropy when conditioned on the messages from the verifier. This rules out some natural tests of quantumness. For example, any computational task in NP that is considered beyond the reach of polynomial-time classical computation, such as factoring or discrete-logarithms, suffices as a test of quantumness; however, \old{it does not}\rev{not all such tasks} generate any new randomness since the correct output is \rev{often} completely determined by the input. 

In practice, one can ensure that the quantum generator can only perform computations scaling polynomially in some security parameter by requiring that responses from the prover are returned under some chosen time threshold. However, it is not natural to place restrictions on the amount of pre-computation that can be performed by the prover (or other adversaries) before the beginning of the protocol. Consequently, a protocol for certified randomness typically includes random messages from the verifier, since otherwise an adversarial prover could pre-compute a list of acceptable responses to the verifier's deterministic messages. Moreover, a protocol may also \old{require}\rev{use} some internal (private) randomness\old{independent from the random messages}, such as determining the subset of rounds to be used for testing \cite{aaronson2023certified} without allowing the adversary to predict the untested rounds. \rev{The random messages as well as the private randomness may be perfectly random (near-uniform) or imperfectly random (far from uniform). In the perfectly random case, the goal of the protocol is to produce output randomness that is longer than the input randomness, i.e., \emph{randomness expansion}. In the imperfectly random case, the goal is to output near-perfect randomness, i.e., \emph{randomness amplification}. Since protocols for certified randomness based on computational assumptions discussed in this Perspective are only analyzed assuming perfectly random inputs, the security of the protocols is only known to hold in the randomness expansion setting. However, we believe it can be fruitful to analyze the security of similar constructions for randomness amplification.}

\rev{Additionally, as discussed before, such an interactive protocol based on computational assumptions does not require physical assumptions about the quantum device and the randomness can be verified even if the device is accessed remotely. This \emph{remote verification} property is highly desirable, both due to the ease of use and the potential to establish consensus among multiple parties.}

\rev{Finally, since the quantum processes used in certified random protocols are \emph{fundamentally random} instead of practically random due to the adversary's lack of information, the resulting outputs satisfy \emph{everlasting security}. This is to say that for an output to be secure, the various assumptions about the adversary only need to hold during protocol execution. This is because unlike any classical process, the quantum protocols do not have information hidden from the adversary that could be later learned to deterministically predict the output. This holds both for any given output, and when an adversary may have seen previous outputs and then would like to predict subsequent outputs.}

\rev{We summarize these properties of certified randomness in Box 1. Additionally, we summarize current certified randomness protocols based on computational assumptions in Box 2. Finally, for a comparison of these protocols, refer to Table \ref{tab:comparison}. We additionally highlight the protocol by Aaronson and Hung \cite{aaronson2023certified}, which has been refined and experimentally implemented \cite{jpmc_cr}, making it a highly promising early candidate to realize industrial value from quantum computing. 
However, the fidelity, speed, and reliability of the quantum computers will need to improve to make the protocol more practical. Furthermore, this protocol requires expensive (exponential-time) classical verification. Other protocols will likely remain out of reach for near-term quantum hardware since they require execution of deep quantum circuits, but have the benefit of efficient verification compared to the protocol by Aaronson and Hung.}

\begin{figure*}[ht!] %
\centering
\begin{minipage}{0.95\textwidth}
\RaggedRight %
\begin{mybox}[Box 1: Certified Randomness]
\setlength{\parskip}{0.5em}
\setlength{\parindent}{1em} %

\textbf{Interaction:} Consider an interactive protocol between a classical verifier $V$ and a quantum prover $P$. $V$ consumes random bitstrings $C_1,\dots,C_L$ for an $L$-round protocol. In the $i$th round, $V$ sends $C_i$ to $P$, and $P$ returns a response $R_i$ by performing some computational task prescribed by $C_i$. Denote $\mathbf{C}=C_1\otimes\dots\otimes C_L$ and $\mathbf{R}=R_1\otimes\dots\otimes R_L$, $\mathbf{C}$ has $n_{\mathbf{C}}$ bits, and $\mathbf{R}$ has $n_{\mathbf{R}}$ bits. Additionally, $V$ may hold some internal (private) randomness $C_{\rm p}$ of $n_{\rm p}$ bits and $S$ of $n_S$ bits which are never sent to $P$.

\textbf{Verification:} A verification function is a function $\mathrm{Ver}(C_{\rm p},\mathbf{C},\mathbf{R})\in\{0,1\}$ that outputs whether the prover passes verification.

\textbf{Soundness:} A randomness extractor is a function $\mathrm{Ext}(S,\mathbf{R})=O\in\{0,1\}^{n_{\rm out}}$ with $n_{\rm out}<n_{\mathbf{R}}$. A certified randomness protocol is sound if either $\mathrm{Ver}(C_{\rm p},\mathbf{C},\mathbf{R})=0$ with high probability over $\mathbf{R}$ and $C_{\rm p}$, or $O$ is almost uniformly distributed conditioned on the side information.

\textbf{Completeness:} There exists a polynomial-time quantum prover that outputs $\mathbf{R}$ such that $\mathrm{Ver}(C_{\rm p},\mathbf{C},\mathbf{R})=1$ with high probability.

\textbf{Useful properties:}
\begin{enumerate}
    \item Everlasting security: Even if the assumptions required for the security of the protocols no longer hold in the future, the outputs generated in the past are still nearly perfectly random conditioned on the side information.
    \item Classical remote verification: Verification is possible even if the interaction with the prover is over the internet with only classical communication.
    \item Randomness expansion: producing a nearly perfectly random output longer than the random input consumed by the protocol (i.e. $n_{\rm out}>n_{\mathbf{C}}+n_{\rm p}+n_S$).
    \item Randomness amplification: producing a nearly perfectly random output given imperfectly random input.
\end{enumerate}

\centering
    \includegraphics[width=\linewidth]{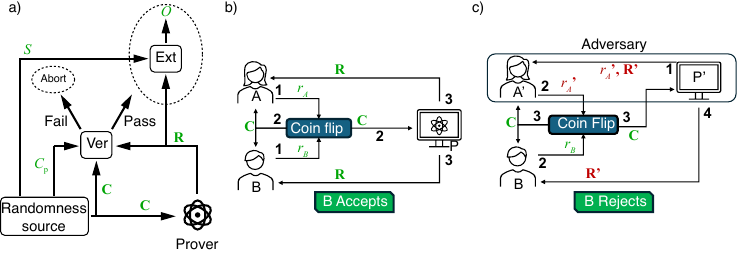}
    \caption{(a) Schematic for the interactive protocol of certified randomness. (b, c) Schematic for jointly certified randomness in the honest and malicious cases. Here, $r_A,r_B,r'_A$ are input randomness for the coin flipping protocol sent from the participants. In (b), Alice ($A$) and Bob ($B$) both honestly engage in a coin-flipping protocol to produce a challenge, which the quantum server ($Q$) honestly responds to. Because the response is honestly prepared for the challenge that was verifiably produced by the coin-flipping, Bob (as well as Alice) accepts. In (c), a malicious Alice ($A’$) and server ($Q’$) collude to attempt to influence Bob to accept pre-determined responses by tailoring Alice’s inputs based on the desired response. However, the coin flipping protocol guarantees that Alice cannot collude with the server to bias challenge from uniformly random as long as Bob is honest. Therefore, Bob is able to detect the pre-determined responses by verifying that they do not correspond to the challenge and reject.}
    \label{fig:joint-and-not}

\end{mybox}
\end{minipage}
\end{figure*}

\begin{table*}[ht!] %
\centering
\begin{minipage}{0.95\textwidth}
\begin{mybox}[Box 2: Protocols for Generating Certified Randomness with Computational Verification]
\setlength{\parskip}{0.5em}
\setlength{\parindent}{1em} %

Randomness can be certified by proving that a sequence of bits originated from a
random process that a deterministic algorithm cannot replicate. Drawing samples from the output distribution induced by a random quantum circuit, also known as Random Circuit Sampling (RCS), is a task believed to be intractable for a classical computer \cite{aaronson2017complexity, Aaronson2020, Bouland2018}. Most experimental demonstrations of quantum computational advantage to date used RCS
\cite{Arute2019, Zhu2022, morvan2023phase, qntm_rcs, gao2025establishing}. The bitstrings drawn from a noisy quantum computer score high on the cross-entropy benchmarking (XEB) score, which is hard to spoof classically under certain complexity-theoretic assumptions~\cite{Aaronson2020, Bouland2018}. This makes a high XEB score an effective proof of ``quantumness''

This observation has motivated theoretical analyses showing that samples which achieve a high XEB score must necessarily contain some entropy, provided that the prover device (classical or quantum) generating the samples is bounded by polynomial space and time \cite{aaronson2023certified, bassirian2021certified, jpmc_cr}. Specifically, based on a plausible complexity-theoretic assumption \rev{(the Long List Hardness Assumption (LLHA) \cite{aaronson2023certified,aaronson2023certifiedArxiv}, a custom assumption)}
it was shown that any prover with an XEB score greater than a threshold with a sufficiently high probability must generate $\Omega(n)$ bits of entropy, where $n$ is the number of qubits. \rev{Alternatively, the protocol can be shown to be unconditionally secure relative to a Haar random unitary oracle.} Recently, a refined and strengthened version of the simplified protocol for certified randomness proposed in Ref.~\cite{aaronson2023certified} was realized experimentally using a trapped-ion quantum computer and exascale classical supercomputers \cite{jpmc_cr}. 

The security of the protocol discussed above derives from the challenge circuits being hard to simulate. However, verification by computing the XEB score requires exponential computational resources. It is therefore desirable to have a protocol which can be verified efficiently. 

The first protocol for a single-device certified randomness was proposed by Brakerski et al.~\cite{brakerski2021cryptographic} and later optimized by Mahadev, Vazirani and Vidick~\cite{mahadev2022efficient}. These protocols admit classical polynomial time verification and their security is based on the quantum hardness of the Learning with Errors (LWE) problem, which is a common cryptographic assumption used in post-quantum cryptography and quantum fully homomorphic encryption. The verifier has to generate a trapdoor that is kept secret from the prover. \old{The messages exchanged between the verifier and the prover relate to arithmetic over the finite field $\mathbb{Z}_q$. Against a computationally bounded prover, it is highly likely that the min-entropy of verifier-accepted outputs scales like $n - \ell \log q$, where $\ell$ is a parameter of the LWE problem.}Since an honest prover is required to be able to coherently perform modular arithmetic, this protocol is not possible \old{with near-term}\rev{on today's} quantum hardware.
 
Yamakawa and Zhandry~\cite{yamakawa2022verifiable} present yet another candidate protocol for certifiable randomness\old{, whose}\rev{. Its} correctness is based on the Aaronson-Ambainis \rev{(AA)} conjecture, which is a widely believed mathematical conjecture in boolean analysis~\cite{aaronson2014need}. The protocol is based on an NP-search problem constructed from an error correcting code. In Ref.~\cite{yamakawa2022verifiable}, this problem was demonstrated  to be exponentially hard for a classical probabilistic computer but solvable by a quantum computer in polynomial time, in the random oracle setting. In the protocol, the verifier poses a search problem to the prover, with the random oracle instantiated with a cryptographic hash such as SHA-2, where the problem size $n$ serves as a security parameter. It can be shown that the response of a polynomial quantum prover answering the search problem must have entropy that grows with $n$. An honest prover in this protocol needs to perform a decoding operation for an error correcting code, which is not feasible on \old{a near-term}\rev{today's} quantum computers.

\old{The primary downside of quantum random number generation protocols based on non-local games is the lack of remote verification, i.e., a remote user with access to bits generated from such an experiment has no way to independently verify the non-signaling assumption. Recently, Kalai et al.~\cite{kalai2023quantum} have demonstrated that a non-local game can be compiled into a single-prover interactive proof of quantum advantage, assuming the existence of a sufficiently powerful quantum homomorphic encryption scheme~[brakerski2018quantum,mahadev2020classical]. It is therefore reasonable to expect that certified random number generation protocols based on non-local games can be similarly compiled to a single prover protocol based only on computational assumptions; however, the details of such a protocol remain to be fully worked out.}

\centering
        \begin{tabular}{|p{3.7cm}|p{2cm}|p{2.5cm}|p{1.7cm}|p{3.5cm}|}
        \hline
         Protocol & Non-local games & Aaronson and Hung \cite{aaronson2023certified} & Brakerski {\sl et al.} \cite{brakerski2021cryptographic} & Yamakawa and Zhandry \cite{yamakawa2022verifiable} \\
         \hline
         Security model and conditions & \cellcolor{green} Unconditional & \cellcolor{red} LLHA or a Haar random oracle & \cellcolor{yellow} Hardness of LWE & \cellcolor{yellow} AA conjecture and the random oracle model \\
         \hline
         Randomness expansion & \cellcolor{green} Yes & \cellcolor{green} Yes & \cellcolor{green} Yes & \cellcolor{green} Yes \\
         \hline
         Randomness amplification & \cellcolor{green} Yes & \cellcolor{yellow} Not proven & \cellcolor{yellow} Not proven & \cellcolor{yellow} Not proven \\
         \hline
         Verification complexity & \cellcolor{green} Constant & \cellcolor{red} Exponential & \cellcolor{green} Polynomial & \cellcolor{green} Polynomial \\
         \hline
         Remote verification & \cellcolor{red} No & \cellcolor{green} Yes & \cellcolor{green} Yes & \cellcolor{green} Yes \\
         \hline
         \rev{Realized experimentally} & \cellcolor{green} Yes\rev{~\cite{bell2004speakable,brunner2014bell,mironowicz2024generalized,joarder2022loophole,nath2024single}} & \cellcolor{green} Yes\rev{~\cite{jpmc_cr}} & \cellcolor{red} No & \cellcolor{red} No \\
     \hline
\end{tabular}
    \caption{Comparison between different protocols for certified randomness generation.
    }
    \label{tab:comparison}

\end{mybox}
\end{minipage}
\end{table*}

\begin{table*}
    
\end{table*}

\subsection{Jointly certified randomness}

In some applications, multiple parties may need to certify randomness \emph{jointly}. For instance, in a lottery, digital election, or census, a specific verifier may be incentivized to collude with the quantum generator, in order to benefit from the appearance of using a random seed while actually using one that is predictable. In such cases, the parties may want to generate randomness that can be certified even if a subset of them colludes with the generator to trick the other (honest) parties into thinking that the generated string has more entropy than it does in reality. \rev{With the remote verification property of certified randomness, this becomes possible.}

Interactive certified randomness protocols can be straightforwardly extended to realize jointly certifiable randomness.
For multiple verifiers to obtain the same random string, they must present the same challenge to the generator. The main challenge is that if one or more verifiers is untrusted, they may attempt to disproportionately bias the challenge generation process towards those challenges that are easier for the generator. For example, the generator may pre-compute deterministic responses to a subset of possible challenges. This bias invalidates the central assumptions that are needed for the security of the certified randomness protocol, which require that the parties share the bits used for computing the challenges, and the shared bits are near uniformly random.

To overcome this problem and jointly construct challenges, we can use distributed protocols that allow parties to jointly generate random strings where the parties have access to a source of private randomness. These have been extensively studied in the cryptography literature, beginning with the seminal work of Blum~\cite{blum1983coin}, and are often referred to as protocols for \emph{Coin Flipping}.\footnote{Unlike the quantum jointly certified randomness protocols, these classical coin flipping protocols can not achieve \emph{randomness expansion}, i.e., the randomness generated jointly is no larger than the combined inputs of the parties.} A concrete coin flipping protocol can be obtained using the secure multi party computation protocols of Ben-Or, Goldwasser, and Wigderson~\cite{wigderson1988completeness}. \old{Combined with a coin flipping protocol to agree on a common random seed to be used to create the challenges, interactive certified randomness protocols enable joint certification of randomness as outlined in \Cref{fig:joint-and-not}.}\rev{The participants can use the coin flipping protocol to generate challenges for use in the certified randomness protocol, as well as other inputs (e.g., the seed to extract the output). 
The randomness expansion property guarantees that the output is a longer sequence of jointly certified uniform bits. In \Cref{fig:joint-and-not}, we depict the jointly certifiable protocol in the honest and the malicious setting, demonstrating the inability for a participant to successfully collude with the generator and bias the output.}

The intuition behind the security of such a construction is straightforward. The security of the underlying single verifier certified randomness protocol relies on the fact that the challenges being sent to the certified randomness generator are unpredictable to the generator. In the single-verifier setting, this is guaranteed by the client's assumed honesty. In the multi-verifier setting, using a coin flipping protocol to determine the challenges allows honest parties to guarantee that fresh randomness is used to prepare the challenges, preventing collusion between dishonest parties and the generator.

\section{Applications}\label{sec:applications}

We now present the proposed applications. For each of the domains, we focus on applications with the clearest potential to be impacted by the ability to certify randomness, giving intuition for the impact and sketching out the constructions in each case.

\subsection{Cryptographic Primitives}

\label{sec:cryptography}

Just as cryptography is integral to the secure operation of today’s global information technologies, randomness is integral to the secure operation of that same cryptography. A symmetric key encryption scheme, for example, transforms a random key and a message into a ciphertext from which an adversary can learn nothing about the message. However, if the randomness used to produce the keys underlying these and other secure cryptographic tools was predictable to adversaries, an attacker would be able to bypass the securely constructed schemes altogether by simply guessing the key. More generally, it is 
known that large swaths of cryptography are impossible to securely realize using biased sources of randomness \cite{dodis2004possibility}. 

At the same time, true randomness is scarce in classical systems. Systems may try to offer randomness gathered from various phenomena that are believed to be unpredictable, such as patterns of memory access and key or mouse interactions, but these sources may fail to provide the anticipated
entropy \cite{heninger2012mining}. Perhaps more worryingly, even if the sources used are unpredictable in principle, in general one cannot guarantee that their output is not correlated with some outside system, or, for example, that the bits available were not pre-generated from the source and shared with an attacker prior to use, as in the case of the so-called ``memory stick'' attacks \cite{acin2016certified}.

\subsubsection{Public Randomness Beacons}
\label{sec:beacon}

\rev{We now show how to create a public randomness beacon that broadcasts a large amount of randomness using jointly certified randomness expansion.} %
First proposed by Rabin \cite{rabin1983transaction}, the ideal functionality of a public randomness beacon proposes to capture the availability of a trusted third party that can produce and distribute fresh randomness to parties at regular intervals, with the guarantees that no parties have the ability to manipulate or predict the randomness emitted. While there may be multiple alternative definitions of these guarantees, we use the NIST reference for randomness beacons \cite{kelsey2019reference} as the baseline.  Publicly available randomness is a useful tool for a variety of applications, including electronic voting, blockchain, lotteries, as well as for many cryptographic protocols. This valuable primitive may be enhanced using certified randomness.

Intuitively, the major guarantees of a randomness beacon are that the outputs are unpredictable to outsiders prior to their publication, and that outputs are fresh -- i.e., that the randomness used to generate them was not sampled long in advance of the output. However, if the beacon's randomness sources are compromised and the outputs are predictable to some external party, or otherwise fail to provide the stated amounts of entropy, the unpredictability of the outputs will suffer. Likewise, a compromised beacon may choose to sample from the sources in advance, damaging the freshness offered. 

Jointly certifiable randomness expansion may be useful in constructing randomness beacons that achieve the desired guarantees of unpredictability and freshness without requiring the honesty of the beacon operator.
We note that the use of certified randomness for similar purposes has been discussed previously, for example in \cite{brandao2020notes}. \rev{More recently, a randomness beacon was proposed that allows the user to certify the generated randomness in a device-independent fashion as long as the set up is trusted~\cite{kavuri2024traceable}.}

Assuming some public-key infrastructure and quantum-safe digital signature schemes \cite{ducas2018crystals, bernstein2019sphincs+}, a collection of verifiers can engage in the jointly certified randomness expansion protocol to generate certifiably random outputs and provide them to the beacon to be used as a source of randomness for the beacon. In addition to the output, the verifiers can provide digitally signed copies of the transcript and output to be emitted alongside the randomness of the beacon. These signed objects allow consumers of the beacon to verify that a sufficient quantity of verifiers, e.g., a majority of the verification network, certify that the jointly certified randomness scheme was properly executed, thereby decentralizing the trust away from the beacon and to a larger network. \old{More recently, a randomness beacon was proposed that allows the user to certify the generated randomness in a device-independent fashion as long as the set up is trusted~\cite{kavuri2024traceable}.}

\old{Jointly certified randomness expansion protocols based on quantum advantage may be useful in reducing the trust assumption from the generation process. }However, the construction sketched above comes at the notable disadvantage of requiring interaction by the verifiers prior to the output of every block, which may be untenable in certain contexts, for example, when the set of verifiers is large. To mitigate this issue, verifiable delay functions (VDFs)~\cite{boneh2018verifiable} may be used to reduce the frequency with which verifiers must inject fresh entropy into the beacon. Informally, a VDF guarantees that, even with some bounded pre-processing time, no adversary can generate the output in parallel time meaningfully faster than some fixed value for a fixed number of processors. Further, the output can be verified to be correct.

Ultimately, it is desirable for the beacon to be able to use previously output certifiably random blocks to bootstrap the generation of subsequent blocks, which requires uniformly random strings for the seed of a randomness extractor. This seed must not be known by the generator since prior knowledge of the extractor seed may allow the generator to bias the raw output to be correlated with the extractor seed in some way. To avoid this, the extractor seed can be taken to be the output of a VDF on randomness from the previous block, which ensures that the seed is not known to the generator until after it has already publicly output the entropy. In practice, however, a beacon operator may feasibly attempt to circumvent this by exploiting looseness in the timing of certain sub-processes in the block generation. Formalizing and rigorously analyzing a public certified randomness beacon is a promising research area.

\subsubsection{Non-interactive zero knowledge proofs}\label{sec:nizk}

\rev{We may further extend jointly certified randomness expansion to improve the feasibility of non-interactive zero knowledge proofs with limited private randomness and large proof.}
Since their inception in the 1985 seminal work of Goldwasser, Micali and Rackoff \cite{goldwasser2019knowledge}, zero-knowledge (ZK) proofs have attracted notable attention, leading to the development of many real-world applications in areas such as decentralized identity \cite{yang2020zero}, privacy-preserving transactions via blockchain \cite{sun2021survey}, secure and verifiable electronic voting \cite{groth2005non}, Internet of Things \cite{walshe2019non}, supply chains \cite{sahai2020enabling}, and many others \cite{mohr2007survey, morais2019survey}. Informally, a ZK proof system has a prover that aims to convince a verifier about a claim (such as possession of an access token) without revealing any information other than the truthfulness of the claim (such as information about the access token).

While secure interactive ZK proofs are known to exist without requiring any trust assumptions when parties are allowed setup in the form of \emph{shared randomness} \cite{goldreich2004foundations}, such proofs are practically infeasible where one desires to build ZK proofs with either a small constant number of rounds, or, at best, a single interaction from the prover to the verifier \cite{barak2006lower}. This has led to the development of secure non-interactive zero knowledge (NIZK) proofs \cite{rackoff1991non, blum1991noninteractive}. NIZK differs from its interactive counterpart in the fact that one allows only a single communication from the prover to the verifier. For non-trivial problems in the NP complexity class, such proofs cannot exist without requiring additional trust assumptions on the protocol \cite{goldreich1994definitions}. This begs a natural question: what are the minimal trust assumptions required to prove the existence of NIZK proofs which are practically efficient and also come with strong security guarantees? The 
proposed trust-based models for NIZK include 
 the common reference string (CRS) model \cite{blum1991noninteractive}, NIZK with preprocessing \cite{de1990non}, secret-key based NIZK \cite{cramer2004secret}, and the multi-string model \cite{groth2014cryptography} , which all serve to generate shared randomness between the prover and the verifier prior to the execution of the protocol. Due to the practicality benefits and guarantees of universal composability, CRS is the most widely adopted trust model to build commercial applications based on NIZKs \cite{damgaard2000efficient}.

A CRS model assumes that the participating parties executing the protocol have prior access to a shared random string that is guaranteed to be taken from some pre-defined distribution, and that no side information on that string is known to either of the parties. An immediate issue facing the real world applications relying on NIZK proofs based on the CRS model is \emph{how does one ensure that the shared random string has been generated in a trusted manner?} More concretely, how can one ensure that the third party generating the string has not inserted a secret trapdoor in the generation process and thus compromised the 
security of
the
proof? This line of attack, also called \emph{parameter subversion} of CRS, has been studied extensively in the literature~\cite{bellare2016nizks, ananth2021towards, ananth2024nizks}. A prover colluding with the third party subverter could then generate false proofs which would pass the verification test. Alternatively, if the potentially malicious verifier colludes with the subverter, it could allow the secret extraction from the proof of the honest prover, thus breaking the zero-knowledge property.

The standard approach to overcoming the aforementioned issue requires the multiple participating authorities to execute the interactive coin flipping protocol. The authorities first generate a common random string with some bounded bias by requiring the exchange of uniformly random bits during the interaction stage. Next, a technique such as inverse transform sampling or acceptance-rejection sampling is used to transform this string from the uniformly random distribution to the CRS for the target distribution with the same bias \cite{westfall2013understanding, groth2014cryptography, abram2023security}. However, for NIZK proofs, the length of the required string often scales polynomially with the proof size \cite{groth2006perfect}. Thus, for proofs of large size, the above proposal is restrictive as it requires each authority to have $poly(|\textsf{proof size}|)$ bits of private randomness to begin with. 

Since certified randomness allows for polynomial expansion in the randomness of the input string used to produce the CRS, it may be used to enable polynomial (in the proof size) reduction in the amount of private randomness that the participating authorities must provide. Using the jointly \old{verifiable }certified randomness \rev{expansion} scheme, the parties only require small amounts of private randomness to perform the coin flipping step and generate the common seed. Subsequently, the certified randomness expansion protocol is run to generate the longer common random string which then gets transformed into the CRS used for NIZK. The requirement for the two parties to directly engage in a jointly certified randomness protocol can be lessened by using the certified randomness beacon discussed in Sec.~\ref{sec:beacon}, at the cost of introducing some trust assumptions necessary for the beacon.

\subsubsection{Immunization of Pseudorandom Generators}

Given the scarcity of randomness, it is common to use the output of cryptographically secure pseudorandom number generators (PRNGs) seeded on some small starting seed, where well designed PRNGs offer the guarantee that their outputs should be indistinguishable from uniform randomness for all efficient (i.e., polynomial-time) attackers. The subversion of the guarantees of pseudorandom number generators can be due to various reasons \cite{cohney2018practical,Markowsky2014,davis2024possibility} and \rev{can} have severe impacts on security \cite{bernstein2016dual,cohney2018practical,Markowsky2014,NISTValidation}.

The most famous example of such a subversion is presented by Dual\_EC\_DRBG~\cite{bernstein2016dual}. In 2007, NIST, followed by ANSI and ISO, standardized a construction based on elliptic curves, Dual\_EC\_DRBG, as an approved option for cryptographically secure random number generators \cite{NISTSP800-90Rev.1}. By 2007, it was widely noted \cite{rsa_backdoor} that the construction of Dual\_EC\_DRBG allowed for the existence of backdoor information that would allow the holder of said information to learn the internal state of the PRNG after viewing a small amount of its outputs, and therefore predict later outputs of the PRNG. In spite of this potential vulnerability, Dual\_EC\_DRBG remained as one of four standardized cryptographically secure PRNGs in NIST SP 800-90A. In 2013, Snowden revelations confirmed that the backdoor was purposefully installed by the NSA~\cite{NYTSnownedDUALEC}. In 2014, Dual\_EC\_DRBG was withdrawn from the standard.

Dodis et al. \cite{dodis2015formal} initiated the study of methods for \textit{immunizing} PRNGs from backdoors. Informally, an immunizer is a method for post-processing the output of any backdoored PRNG to retain the security guarantees of a secure PRNG even against attackers in possession of the backdoor. Among a number of results in this area, it is known that there cannot exist deterministic immunizers that can operate on the output of a single PRNG \cite{dodis2015formal}.

Certified randomness may be used as an immunizer for potentially backdoored PRNGs. We consider the setting suitable for extracting private randomness from a certified randomness scheme, in which the generator is potentially entangled with an adversary and may act maliciously, but cannot communicate with the outside world. We further require that the generator is not in possession of the backdoor to the PRNG in question. Consider a certifiable randomness scheme in which the input randomness is replaced by a pseudorandom string generated by a potentially backdoored PRNG. From the perspective of the generator, which knows no backdoor, the string is indistinguishable from a truly random string, and so the guarantee of the certified randomness protocol still holds: the output string is uniformly distributed and uncorrelated from the input randomness and any external systems. Furthermore, an external adversary in possession of a backdoor to the PRNG who sees the output, or even the internal randomness, has no advantage in distinguishing future outputs from the uniform distribution, or predicting future outputs. Indeed, this should hold even though the external adversary, with knowledge of the backdoor and the challenges, can predict future challenges to the generator. A summary is given in \Cref{fig:immunization_and_dp} (a).

\rev{For completeness, we further note that a similar argument can be used to mitigate the risk of using poor sources of randomness, in which a hardware RNG fails to satisfy the promised guarantees for various reasons~\cite{Garfinkel2020, Shrimpton2015}.
The case when an RNG provides near-uniform randomness but with some backdoor accessible to an adversary \cite{blaze2011key} is similar to the immunization of backdoored PRNGs above. The case where the randomness is not near-uniform \cite{Shrimpton2015}, however, differs slightly, and it becomes necessary to use a certified randomness protocol that provides randomness amplification as a guarantee, turning a source that is not uniform (but still has sufficiently high entropy) into one that is nearly so. This feature, albeit acknowledged as desirable, has been absent from existing certified randomness analyses based on computational assumptions.}

\subsection{Bias, Fairness, and Privacy}\label{sec:bias_fairness_privacy}

Several applications in data science, machine learning, economics, and mechanism design such as electronic voting \cite{chawla2015power}, privacy-preserving data mining \cite{evfimievski2002randomization}, privacy preserving machine-learning \cite{wang2020privacy}, and private distributed computing~\cite{cramer2015secure} rely on randomness in order to ensure fairness and/or privacy. It may be desirable to preserve these guarantees against an adversary that it is computationally unbounded, especially for those applications where the datasets may remain accessible for long and/or unpredictable amounts of time. This stronger notion, known as ``everlasting privacy" is particularly natural in the setting of electronic voting, where it was first formalized and introduced~\cite{moran2006receipt, arapinis2013practical, haines2023sok}.

\subsubsection{Electronic Voting}

Principled electronic voting systems strive for \emph{verifiability}, which allows external validators to verify whether the election results accurately reflect voter choices (even in the presence of a certain number of malicious parties in the voting and tallying protocols), and \emph{privacy}, which ensures that the data exposed by the protocol does not reveal more about an individual voter's choices than the election result itself. Many well-known protocols for electronic voting such as Helios~\cite{adida2008helios}, Civitas~\cite{clarkson2008civitas}, Belenios~\cite{cortier2019belenios}, Ordinos~\cite{kusters2020ordinos}, etc., rely on cryptographic constructions where voters encrypt their votes and publish the cyphertexts based on a public key, and the cyphertexts are subsequently decoded by talliers using a secret key. The security of these protocols was based on computational assumptions such as Diffie-Hellman~\cite{escala2017algebraic}. As a consequence, these protocols do not offer everlasting privacy against future adversaries with access to the public transcripts of previous elections and more powerful algorithms or computers. \old{There}\rev{Protocols} have since been \old{protocols~\cite{iovino2017using, querejeta2020netvote}} proposed that offer everlasting privacy, but make use of random seeds\rev{~\cite{iovino2017using, querejeta2020netvote}}. We note that if these seeds are instead pseudorandom, then the protocol is potentially vulnerable to an unbounded adversary with access to a long stream of random seeds used by the protocol. We propose therefore that the use of genuine certifiable randomness in place of pseudorandomness offers a notable uplift for privacy protocols. As the specific manner in which pseudorandomness is used in e-voting protocols differs from protocol to protocol, so does the degree to which a compromise of PRNG can be tolerated. We now illustrate some specific vulnerabilities, and a potential uplift from certified randomness, in the context of differential privacy.

\subsubsection{Differential Privacy}
\label{sec:differential-privacy}

\rev{Everlasting security afforded by certified randomness can uplift differential privacy pipelines and obtain everlasting privacy.}
The potential requirement for everlasting privacy guarantees arises in many areas of data science. To illustrate PRNG-based attacks on these systems, and the corresponding uplift from the use of certified randomness, we focus our discussion on the well-studied notion of ``differential privacy", where the role of pseudo-randomness in popular protocols is similar, and theoretically well-understood.

The goal of \emph{privacy-preserving} data analysis is to enable the computation of useful statistics over a population without compromising the privacy of individual entries. The data is usually collected by a trustworthy curator and may be revealed in two settings. In the non-interactive setting, the curator publishes a fixed transcript from the data and all further computation is based only on this transcript. In the interactive setting, the curator acts as an intermediary between a database and users seeking to compute a statistic based on it. The users are permitted to issue certain queries, while the curator is permitted to modify its responses to protect the privacy of respondents. There are many approaches to private data analysis from the points of view of statistics, database theory, and cryptography.
A widely accepted and commonly used notion in private data analysis is that of differential privacy (often abbreviated as DP), introduced and formalized by Dwork, McSherry, Nissim, and Smith in 2006~\cite{dwork2006calibrating}. Differential privacy captures roughly the guarantee that the outcome of a statistical analysis carried out on a database is not notably affected by the inclusion or exclusion of any individual entry in the data. It ensures that a respondent incurs inconsequential risk by joining the database, despite the fact that the distribution of the computed statistics may disclose information about the population of respondents in general.

Differential privacy is typically ensured using a randomized noise mechanism, where the data curator adds a random perturbation to the queried data before returning it to the querying user or application (see~\cite{dwork2014algorithmic} for more exposition). It has been demonstrated~\cite{dwork2006calibrating} that the level of noise can be chosen so as to ensure differential privacy while still allowing for algorithmic utility. This discovery has made DP a very successful paradigm with several distinct advantages: originating from its clear mathematical definition, precise trade-offs between privacy and utility, readily composable guarantees, and resistance to attacks based on auxiliary data~\cite{near2023guidelines}. It has also been argued that differential privacy may be the only concrete definition of privacy that satisfies regulations such as FERPA \cite{nissim2017bridging} and GDPR \cite{cummings2018role}, while many common practices in modern machine learning fall short. It is increasingly accepted as a standard for privacy~\cite{near2023guidelines} and has recently been deployed for the US Census of 2020~\cite{ruggles2019differential, Garfinkel2020}.

Randomized mechanisms for differential privacy, such as the ones listed above, can be inverted if the random noise added to queries is predictable. Additionally, it is often possible for an external unbounded adversary to obtain samples \rev{of the added random noise}\old{from the randomness generator used by the protocol}. For instance, the adversary could masquerade as a user\rev{,} \old{and }contribute to the dataset, \old{after which it} \rev{and} make\old{s} repeated queries to which it knows the true response (before randomization). As a result, these protocols do not exhibit everlasting privacy if a pseudorandom generator is used to randomize the response to each query\old{.} \rev{since the adversary in this setting is
unbounded, enabling it to break
pseudorandomness 
even 
in the absence of a
a backdoor.}
We argue that \old{using }a certified quantum randomness generation protocol may be \old{able}\rev{used} to immunize a protocol against a subset of such attacks. 

\begin{figure}
    \centering
    \includegraphics[width=\columnwidth]{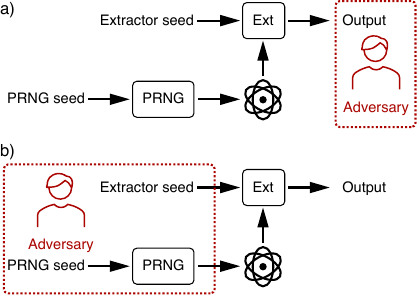}
    \caption{The dashed boxes represent information the adversary can access but not control. (a) Schematic for the security model of PRNG immunization. The adversary can access the protocol output. However, because the certified randomness protocol guarantees the output is independent of PRNG outputs, the adversary cannot predict future outputs. (b) Schematic for the security model of differential privacy. At a later date, an unbounded adversary can learn the PRNG outputs and the PRNG seed. However, because the certified randomness protocol guarantees the output is independent of PRNG outputs, the adversary cannot learn the protocol output. This holds even if the adversary knows the extractor seed.}
    \label{fig:immunization_and_dp}
\end{figure}

\old{Clearly, we cannot immunize against all attacks by unbounded adversaries, as the randomness generation protocol can itself be compromised. Specifically, an unbounded adversary can deterministically perform the computational task posed during the randomness generation protocol if it is able to communicate with the quantum generator during the protocol. We therefore consider a more limited adversary model,}\rev{We consider an adversary model} where the adversary is non-communicating with the quantum prover, although they may pre-share quantum entanglement. Additionally, the adversary is given access to an arbitrarily long list of samples drawn from the \rev{pseudorandom source}\old{random noise used by the privacy mechanism (in practice, this may be obtained by making queries with a known response as described above)}. In this model, the \old{prior samples available to the adversary}\rev{pseudorandomness} can be naturally modeled as the side information available to the adversary. \old{If the computational assumptions of the definition hold during the certified randomness protocol, the output of the certified randomness protocol, if accepted, contains new bits of randomness that are independent of the side information.}\rev{If the computational assumptions hold during the certified randomness protocol and the output is accepted, then the output contains new bits of randomness that are independent of the side information.} When these bits are used to implement the randomized response for a new query, this query should not reveal more information than that prescribed by the DP protocol, preserving privacy guarantees. A summary is given in \Cref{fig:immunization_and_dp} (b).

\subsubsection{Financial Markets}

\rev{We now discuss how public randomness beacons enabled by certified randomness can enhance the trustworthiness and robustness of some operations in financial markets.}
Beyond the setting of everlasting privacy, protocols employing pseudorandomness are subject to an additional attack if the party implementing a protocol is incentivized to bias its outcomes. Such a party may implement the protocol using a PRNG with a predictable seed, which allows it to undo the fairness or privacy guarantees. Since a PRNG cannot be differentiated from true randomness using a polynomial number of observations of the output bits, such a compromise is challenging to detect by simply monitoring the protocol execution. Furthermore, even if a protocol is not compromised, the potential for manipulation in this manner may lead to a decreased perception of fairness. We illustrate some instances where this is relevant in the context of financial markets.

Randomness can be used to fairly allocate a scarce resource. One example is the share allocation in an oversubscribed initial public offering (IPO). In an IPO, a percentage of the shares offered may be allocated to retail investors, who submit bids committing to buying shares at the final IPO price. If the total number of shares requested in retail bids is higher than the number of shares allocated to retail investors, a lottery determines which bids are filled~\cite{Anagol2018}. Such IPO lotteries are common in some markets such as India~\cite{Anagol2018} and China~\cite{Hu2019}, and have been advocated due to their efficiency and fairness~\cite{Schwartz2016IPO}. As with any lottery, perception of fairness by all parties of the random draw is essential to the legitimacy of the procedure. A public certified randomness beacon discussed in Sec.~\ref{sec:beacon} could be used to affirm to all parties that the lottery has been executed fairly and that the result could not have been manipulated.

Randomness can also be used to improve the perceived fairness of markets. Concerns have been raised regarding the possibility that below a certain time threshold, high-frequency traders earn their profits at the expense of slower market participants without providing market utility like liquidity~\cite{ORIOL2024,Marcus2024}. To tackle these concerns, some exchanges have disincentivized high-frequency trading by introducing randomized or fixed delays between the receipt of a trade and its execution (see Ref.~\cite{Khapko2021} for a review of implementations). A certified randomness beacon may be used improve the transparency and fairness of order execution, which is highly sensitive due to its potential to impact the ability of market participants to make profits. However, an important engineering challenge is the integration of the certified randomness protocols into the infrastructure of exchanges, which often operate on sub-millisecond timescales.

\subsection{Blockchain and Decentralized Applications}

Decentralized applications aim to minimize the concentration of trust within a multi-party setup~\cite{cachin2017blockchain}, and were originally used as ledgers where transactions were recorded in blocks linked together cryptographically~\cite{nakamoto2008bitcoin}. In recent years, progress in smart contracts enabled implementation of a wide range of services on blockchains. Smart contracts are code instructions that manipulate the blockchain state in a validated way. As smart contracts perform increasingly sophisticated services, many of them require randomness to fairly implement their functionalities~\cite{wood2014ethereum,avalanche}. For example, decentralized applications for gaming and market design~\cite{2019web3} require randomness generation to ensure fairness across players and market participants. In addition, decentralized applications often need to perform lotteries, auctions, and random elections~\cite{shi2022integration} -- decisions that may have high financial and political stakes. For such sensitive applications, the integrity of randomness used is of paramount importance. There are two primary ways for blockchain applications to obtain randomness. They can either use randomness natively available on the blockchain (``on-chain'' randomness) or import random bits from an external provider (``off-chain'' randomness). 

The native on-chain randomness may use the block's hash~\cite{nakamoto2008bitcoin}, generated in the mining process from block contents, as its source of unpredictability, or \rev{use} more sophisticated methods such as RANDAO~\cite{wood2014ethereum,avalanche} that accumulate randomness over the history of the blocks. Various public blockchains operate with a ``proof of stake'' mechanism that uses on-chain randomness derived from RANDAO~\cite{wood2014ethereum, avalanche}. However, random numbers generated using blockchain contents alone 
present security concerns since blockchain actors and operators often have access to temporary blocks before the blocks get publicly consolidated into the blockchain, making it possible to anticipate and maliciously influence the generated randomness~\cite{alpturer2024manipulation,amiet2021, report_reroll2022}. \rev{These issues are amplified in small networks with fewer participants. Augmenting RANDAO with public certified randomness beacon of Sec.~\ref{sec:beacon} can mitigate some of these challenges in the context of ``proof of stake'', though more elaborate solutions may be required for other applications as we discuss now.}

The limitations of on-chain randomness have motivated the introduction of randomness from external off-chain services, such as verifiable random functions (VRF)~\cite{micali1999verifiable}. VRF commits to a random public-private key pair by releasing a public key. Whenever a client requests randomness from an off-chain VRF, it invokes the VRF server with a fresh seed input $x$. The VRF server then responds by mixing the seed with the secret key to output $y$, along with a proof $\pi$ of pseudorandomness. Any observer (e.g., a smart contract) in the blockchain can verify the pseudorandomness of $y$ using the proof $\pi$, the public key, and the initial seed $x$. However, this requires trust in unpredictability of the VRF private key, which may not hold in the case of a colluding or compromised VRF provider.

Certified randomness amplification may be an alternative off-chain randomness source that does not require this trust assumption.
A smart contract may use imperfect on-chain randomness to generate challenges $C$ for a certified randomness amplification protocol. The protocol outputs $X$, which the smart contract combines with another imperfect on-chain random input $Y$ using a two-source extractor to produce near-perfectly random $Z=\mathrm{Ext}(X,Y)$. We note that care is required to secure this procedure against manipulation of extractor inputs, namely to ensure that the inputs satisfy the necessary Markov source condition \cite{Foreman2023} and are not dependent on one another. It may be possible to enforce this by temporarily ``hiding'' the inputs $X$ and $Y$ using a verifiable delay function, similarly to our interaction efficient randomness beacon in \Cref{sec:beacon}.

\rev{Furthermore, in practical settings, the output of VRF is expanded using an on-chain pseudorandom number generator~\cite{Chainlinkvrf, Chainlinkvrfcontract}. Certified randomness expansion can be used instead, which may offer better security guarantees in some cases.}

\section{Discussion}

\rev{The applications outlined in this perspective show that certified randomness generated by quantum computers has the potential to add value in a broad range of domains, including cryptography, differential privacy, financial markets, and blockchain. 
For example, the ability to certify randomness can prevent an adversary from predicting private keys by abusing compromised hardware or pseudorandom number generators. More broadly, the ability of multiple jointly distrustful parties to certify that a shared string has been generated randomly without manipulation reduces the communication costs and improves the security of trustless protocols across all the domains discussed.}

\rev{This perspective includes only a small subset of applications which may benefit from certified randomness. Given the ubiquity of randomness as a resource in computer science, we anticipate the identification of many additional applications that benefit from certified randomness expansion, remote verification, and everlasting security.}

\rev{An important limitation of the current state-of-the-art is the lack of rigorous security analyses for many of the applications outlined in our perspective. While we believe they can be formalized, doing so requires careful work that is beyond the scope of this perspective and may be a fruitful direction for future research. Additionally, as highlighted in Ref.~\cite{jpmc_cr}, the protocols implementable on today's quantum computers may not be directly applied to production applications due to high verification cost and limitations of security analysis, among other factors. Further improvements to protocols themselves and the quantum hardware implementing them are required to realize the value of certified randomness hypothesized in this perspective.}

\section*{Acknowledgments}
We thank our colleagues at the Global Technology Applied Research center of JPMorganChase for support and helpful discussions. Special thanks to Giuseppe Di Cera for detailed feedback on the manuscript.

\old{The applications outlined in this perspective show that certified randomness generated by quantum computers has the potential to generate value in a broad range of domains. Realizing this value requires both hardware and theoretical progress to improve the cost and security of certified randomness generation. Furthermore, the security analysis of the applications may need to be formalized. As a classically infeasible primitive realizable on near-term quantum hardware, certified randomness is a promising early application of quantum computers.}

\bibliography{main}

\section*{Disclaimer}
This paper was prepared for informational purposes by the Global Technology Applied Research center of JPMorganChase. This paper is not a product of the Research Department of JPMorganChase or its affiliates. Neither JPMorganChase nor any of its affiliates makes any explicit or implied representation or warranty and none of them accept any liability in connection with this position paper, including, without limitation, with respect to the completeness, accuracy, or reliability of the information contained herein and the potential legal, compliance, tax, or accounting effects thereof. This document is not intended as investment research or investment advice, or as a recommendation, offer, or solicitation for the purchase or sale of any security, financial instrument, financial product or service, or to be used in any way for evaluating the merits of participating in any transaction.

\end{document}